\documentclass[12pt]{article}
\date{}
\title{Electron-Neutrino Bremsstrahlung in Electro-Weak Theory}
\author{Indranath Bhattacharyya\\
Department of Applied Mathematics\\
University of Calcutta, Kolkata-700 009, INDIA\\E-mail :
$i_{-}bhattacharyya@hotmail.com$\\} \vskip .1in
\begin{document}
\maketitle \baselineskip .3in \noindent
\begin{abstract}
The electron-neutrino bremsstrahlung process has been considered
in the framework of electro-weak theory. The scattering cross
section has been calculated in the center of mass frame and
approximated to extreme relativistic as well as non-relativistic
case. The rate of energy-loss via this type of bremsstrahlung
process has been obtained both in non-degenerate and degenerate
region. The effect of this electron-neutrino bremsstrahlung
process in different ranges of temperature and density
characterizing the late stages of stellar evolution has been
discussed. It is found from our study that this bremsstrahlung
process is highly important in the non-degenerate region, although
it might have some significant effect in the extreme relativistic
degenerate region. \vspace{0.5cm}\\\noindent{PACS
:\hspace{0.2cm}}{\it 12.15.-y; 13.85.Hd; 13.85.Lg; 14.60.Lm;
14.70.-e; 97.60.-s}
\end{abstract}
\section{Introduction} It is known that neutrino emission processes have
significant contribution in the late stages of the stellar
evolution; when the core of the stars collapses, the neutrino
emission occurs enormously in the temperature range $> 10^{9}$ K.
In that range the neutrino emission process is important due to
extremely large mean free path of the neutrino. It is believed
that the stellar matter (even under some extreme conditions as in
white dwarves or in neutron stars) is almost transparent to the
neutrinos, such in contrast with its behavior with respect to
photons. Once neutrinos are produced inside the star the rate of
energy loss is higher resulting faster evolution. There are
several processes functioning the major role in the energy loss
from star in the late stage through the emission of
neutrino-antineutrino pair. First time it was pointed out by Gamow
and Schoenberg \cite{Gamow} that neutrino might be emitted from
star via $\beta$ decay, which is termed as Urca process.
Poentecorvo \cite{Pontecorvo} showed the theoretical possibility
of the formation of neutrino pairs in collisions between electrons
and nuclei and the process was investigated with possible
application in astrophysics by Gandel'man and Pinaev
\cite{Gandel'man}. The reaction rate of this process was
calculated in detail by Chiu and his collaborators \cite{Chiu1960,
Chiu1961, Chiu}. Dicus \cite{Dicus1972} drew a brief outline about
some neutrino emission processes from star, though there might
exist a few more processes having remarkable effect during the
later stages of the stellar evolution. Earlier most of the were
studied in the framework of the V-A interaction theory
\cite{Feynman}. Later, photon-neutrino weak coupling theory was
introduced and some of those processes such as neutrino
synchrotron process \cite{Raychaudhuri1970}, photo-coulomb
neutrino process \cite{Raychaudhuri} were studied in this
framework. The advancement of the Standard Model added a new
dimension to study several neutrino emission processes such as
pair annihilation \cite{Dicus1972}, photo production
\cite{Dicus1972}, photon-photon scattering
\cite{Dicus1972,Dicus1993,Abbasabadi1998,Abbasabadi2000},
photo-coulomb neutrino process \cite{Rosenberg, Bhattacharyya}
etc. Recently Itoh et al. \cite{Itoh} reviewed a number of
neutrino emission processes and discussed their significance from
astrophysical point of view. A minor extension of the Standard
Model was done due to the existence of neutrino mass resulted from
the `Solar Neutrino Problem' and `Atmospheric Neutrino Anomaly'
\cite{Santo}. It should be noted that a new theory of weak
interaction is yet to be developed by introducing the neutrino
mass. In the calculations of some weak processes the effect of
this neutrino mass, whatever small it may be, may play an
important role, for example photon-neutrino interaction
\cite{Dicus2000,Dodelson}.\\\indent In this paper we have studied
the `electron-neutrino bremsstrahlung process' given by
$$e^{-}+e^{-}\longrightarrow e^{-}+e^{-}+\nu+\bar{\nu}$$
according to the electro-weak interaction theory. Previously it
was considered by Cazzola and Saggion \cite{Cazzola} while
calculating the energy-loss rate by using Monte Carlo Integration
method without evaluating the scattering cross-section explicitly.
But the scattering cross-section, depending on the energy of the
incoming electron, can give a very clear idea about the nature of
this electro-weak process. It is to be approximated for extreme
relativistic as well as non relativistic limit. Cazzola and
Saggion \cite{Cazzola} considered only the non-degenerate case
though many stars in the later phases such as white dwarves,
neutron stars etc. are degenerate. We cannot ignore the
possibility of occurring the electron-neutrino bremsstrahlung
process in degenerate star. We have considered all such cases
separately and discussed all possible outcomes of this
bremsstrahlung process. We also like to visualize a picture under
what circumstances the process will have some significant effect.
It cannot be denied that due to some approximations a little bit
deviation may occur from the original result, but that will not
deter to predict the physical picture. The role of this process
has been studied thoroughly at different temperature and density
ranges that characterize the late stages of the stellar evolution.
It has also been pointed out in which range this process has
significant effect.
\section{Calculation of scattering cross-section :}
The electron neutrino bremsstrahlung has some structural
similarity with the bremsstrahlung process in quantum
electrodynamics \cite{Vortuba, Wheeler, Joseph}. In this process a
slight complication arises since the identical particles
(electrons) are involved here. It is not possible to identify
which of the two outgoing particles is the `target' particle for a
particular `incident' electron. In classical physics such
identification can be done by tracing out the trajectories. In
quantum physics the two alternatives are completely
indistinguishable and therefore, the two cases may interfere.
There exist 8 possible Feynman diagrams shown in Figure-1 and
Figure-2. The total scattering amplitude for all possible diagrams
can be constructed according to the Feynman rules as follows:
$$\mathcal{M}^{Z}=-\frac{4\pi i e^{2}g^{2}}{8\cos^{2}\theta_{W}
M_{Z}^{2}}[(\mathcal{M}_{1}^{Z}+\mathcal{M}_{2}^{Z}+\mathcal{M}_{3}^{Z}+
\mathcal{M}_{4}^{Z})-(\mathcal{M}_{5}^{Z}+\mathcal{M}_{6}^{Z}+\mathcal{M}_{7}^{Z}+
\mathcal{M}_{8}^{Z})]\eqno{(2.1)}$$
$$\mathcal{M}_{1}^{Z}=[\overline{u}(p'_{1})(C_{V}-C_{A}\gamma_{5})\gamma_{\rho}
\frac{(q^{\tau}\gamma_{\tau}+p_{1}^{'\tau}\gamma_{\tau}+m_{e})}
{(q+p'_{1})^{2}-m_{e}^{2}+i\epsilon}\gamma_{\mu}u(p_{1})]
[\overline{u}(p'_{2})\frac{\gamma^{\mu}}{(p_{2}-p'_{2})^{2}+i\epsilon}u(p_{2})]
[\overline{u}_{\nu}(q_{1})(1-\gamma_{5})\gamma^{\rho}v_{\nu}(q_{2})]\eqno{(2.2)}$$
$$\mathcal{M}_{2}^{Z}=[\overline{u}(p'_{1})\gamma_{\mu}
\frac{(-q^{\tau}\gamma_{\tau}+p_{1}^{\tau}\gamma_{\tau}+m_{e})}
{(q-p_{1})^{2}-m_{e}^{2}+i\epsilon}(C_{V}-C_{A}\gamma_{5})\gamma_{\rho}u(p_{1})]
[\overline{u}(p'_{2})\frac{\gamma^{\mu}}{(p_{2}-p'_{2})^{2}+i\epsilon}u(p_{2})]
[\overline{u}_{\nu}(q_{1})(1-\gamma_{5})\gamma^{\rho}v_{\nu}(q_{2})]\eqno{(2.3)}$$
$$\mathcal{M}_{3}^{Z}=\mathcal{M}_{1}^{Z}(p_{1}\leftrightarrow p_{2},p'_{1}\leftrightarrow
p'_{2})\hspace{2cm}\mathcal{M}_{4}^{Z}=\mathcal{M}_{2}^{Z}(p_{1}\leftrightarrow
p_{2},p'_{1}\leftrightarrow p'_{2})\eqno{(2.4)}$$
$$\mathcal{M}_{5}^{Z}=\mathcal{M}_{1}^{Z}(p'_{1}\leftrightarrow p'_{2})\hspace{0.5cm}
\mathcal{M}_{6}^{Z}=\mathcal{M}_{2}^{Z}(p'_{1}\leftrightarrow
p'_{2})\hspace{0.5cm}
\mathcal{M}_{7}^{Z}=\mathcal{M}_{3}^{Z}(p'_{1}\leftrightarrow
p'_{2})\hspace{0.5cm}
\mathcal{M}_{8}^{Z}=\mathcal{M}_{4}^{Z}(p'_{1}\leftrightarrow
p'_{2})\hspace{1cm}\eqno{(2.5)}$$ where,
$$C_{V}=-\frac{1}{2}+2sin^{2}\theta_{W}\hspace{2cm} C_{A}=-\frac{1}{2}$$
The prefix $Z$ associated with the matrix element and each of its
component indicates that the neutrino anti-neutrino pair emission
takes place through the exchange of $Z$ boson. It is worth noting
that there exist few more diagrams related to the electron
neutrino bremsstrahlung process. For example, there are some
diagrams in which the virtual photon, indicated in the given
figures, might be replaced by $Z$ and $W$ bosons. In addition
Higgs bosons might be present as the virtual lines in some
diagrams. We are discussing this process in view of its effect
during the late stages of the stellar evolution, where,
$$p^{0}_{1}, p^{0}_{2}\ll M_{Z}, M_{W}$$
and so in this energy range all such additional diagrams
containing more than one ofshell gauge boson lines would have
negligible effect compared to the 16 diagrams considered in this
article. We can thus ignore those diagrams to make our
calculations relatively simpler.\\\indent Since the collision
occurs between two identical particles, both the halves of the
phase are identical. In one half of the phase direct scattering
dominates over the exchange graph, whereas in the other half later
one dominates over the former. So it is quite reasonable to
calculate the process in the half of the phase where the graphs
representing the direct scattering dominates. We have carried out
our calculations in CM frame in which
$$\overrightarrow{p_{1}}+\overrightarrow{p_{2}}
=\overrightarrow{p}'_{1}+\overrightarrow{p}'_{2}+
\overrightarrow{q_{1}}+\overrightarrow{q_{2}}=0\eqno{(2.6)}$$
where $\overrightarrow{q_{1}}$ and $\overrightarrow{q_{2}}$ are
the linear momenta of neutrino and anti-neutrino. In this frame we
have to calculate the term $|\mathcal{M}^{Z}|^{2}$ over the spin
sum. It is not very easy task and can be done with some choices
and approximations. We can express the term $\sum
|\mathcal{M}_{1}^{Z}|^{2}$ as
$$\sum |\mathcal{M}_{1}^{Z}|^{2}=X_{\rho\sigma\alpha\beta}(p_{1},p_{2},p'_{1},p'_{2})
Y^{\alpha\beta}(p_{2},p'_{2})N^{\rho\sigma}(q_{1},q_{2})\eqno{(2.7)}$$
$$X_{\rho\sigma\alpha\beta}(p_{1},p_{2},p'_{1},p'_{2})=\frac{1}{4m_{e}^{2}
|(q+p'_{1})^{2}-m_{e}^{2}+i\epsilon|^{2}}[C_{V}^{2}T_{1}-C_{A}^{2}T_{2}
+C_{V}C_{A}T_{3}-C_{V}C_{A}T_{4}]\eqno{(2.8)}$$
$$T_{1}=Tr[(p_{1}^{\tau}\gamma_{\tau}+m_{e})
\gamma_{\alpha}(P^{\tau}\gamma_{\tau}+m_{e})\gamma_{\rho}(p_{1}'^{\tau}\gamma_{\tau}+m_{e})
\gamma_{\sigma}(P^{\tau}\gamma_{\tau}+m_{e})\gamma_{\beta}]\eqno{(2.8a)}$$
$$T_{2}=Tr[(p_{1}^{\tau}\gamma_{\tau}+m_{e})
\gamma_{\alpha}(P^{\tau}\gamma_{\tau}+m_{e})\gamma_{\rho}(-p_{1}'^{\tau}\gamma_{\tau}+m_{e})
\gamma_{\sigma}(P^{\tau}\gamma_{\tau}+m_{e})\gamma_{\beta}]\eqno{(2.8b)}$$
$$T_{3}=Tr[(p_{1}^{\tau}\gamma_{\tau}+m_{e})
\gamma_{\alpha}(P^{\tau}\gamma_{\tau}+m_{e})\gamma_{\rho}(p_{1}'^{\tau}\gamma_{\tau}+m_{e})
\gamma_{5}\gamma_{\sigma}(P^{\tau}\gamma_{\tau}+m_{e})\gamma_{\beta}]\eqno{(2.8c)}$$
$$T_{4}=Tr[(p_{1}^{\tau}\gamma_{\tau}+m_{e})
\gamma_{\alpha}(P^{\tau}\gamma_{\tau}+m_{e})\gamma_{\rho}\gamma_{5}(-p_{1}'^{\tau}
\gamma_{\tau}+m_{e})\gamma_{\sigma}(P^{\tau}\gamma_{\tau}+m_{e})\gamma_{\beta}]\eqno{(2.8d)}$$
$$Y^{\alpha\beta}(p_{2},
p'_{2})=\frac{1}{m_{e}^{2}|(p_{2}-p'_{2})^{2}+i\epsilon|^{2}}
[p_{2}^{\alpha}p_{2}^{'\beta}+p_{2}^{\beta}p_{2}^{'\alpha}+
\{(p_{2}p'_{2})-m_{e}^{2}\}g^{\alpha\beta}]\eqno{(2.9)}$$
$$N^{\rho\sigma}(q_{1}, q_{2})=\frac{2}{m_{\nu}^{2}}
[q_{1}^{\rho}q_{2}^{\sigma}+q_{1}^{\sigma}q_{2}^{\rho}-
(q_{1}q_{2})g^{\rho\sigma}+iq_{1\tau_{1}}q_{2\tau_{2}}
\epsilon^{\tau_{1}\tau_{2}\rho\sigma}]\eqno{(2.10)}$$
$$q=q_{1}+q_{2}=(p_{1}+p_{2})-(p'_{1}+p'_{2})\eqno{(2.11a)}$$
$$P=q+p'_{1}\eqno{(2.11b)}$$
Instead of calculating the term $\sum|\mathcal{M}_{1}^{Z}|^{2}$ it
is more convenient to calculate the term
$$\int\sum|\mathcal{M}_{1}^{Z}|^{2}\frac{d^{3}q_{1}}{2q_{1}^{0}}\frac{d^{3}q_{2}}{2q_{2}^{0}}
\delta^{4}(q-q_{1}-q_{2})$$  Let us consider
$$I^{\rho\sigma}(q)=\frac{2}{m_{\nu}^{2}}\int
[q_{1}^{\rho}q_{2}^{\sigma}+q_{1}^{\sigma}q_{2}^{\rho}-
(q_{1}q_{2})g^{\rho\sigma}+iq_{1\tau_{1}}q_{2\tau_{2}}
\epsilon^{\tau_{1}\tau_{2}\rho\sigma}]
\frac{d^{3}q_{1}}{2q_{1}^{0}}\frac{d^{3}q_{2}}{2q_{2}^{0}}
\delta^{4}(q-q_{1}-q_{2})\eqno{(2.12)}$$
$$=\frac{1}{m_{\nu}^{2}}(Aq^{2}g^{\rho\sigma}+Bq^{\rho}q^{\sigma})$$
It is to be remembered that the neutrino mass is very small
compared to the magnitude of its linear momentum. This is valid
throughout our calculations, even in the non-relativistic case.
Even if the neutrino mass is comparable to the magnitude of the
linear momentum of the neutrino, then no such neutrino
anti-neutrino pair will be emitted and the process becomes
superfluous. This is very much consistent with the Standard Model
which is based on the concept of mass less neutrino. In principle
neutrino may have a little mass, but it is too little to violate
the basic assumption of the well known existing theory. Thus
taking $m_{\nu}\ll q^{0}$ we can evaluate the integral
$I^{\rho\sigma}(q)$ and find the value of $A$ and $B$. It is found
to be
$$A=-B=-\frac{\pi}{3}$$
and also we get
$$\int\sum|\mathcal{M}_{1}^{Z}|^{2}\frac{d^{3}q_{1}}{2q_{1}^{0}}\frac{d^{3}q_{2}}{2q_{2}^{0}}
\delta^{4}(q-q_{1}-q_{2})=X_{\rho\sigma\alpha\beta}(p_{1},p_{2},p'_{1},p'_{2})
Y^{\alpha\beta}(p_{2},p'_{2})I^{\rho\sigma}(q)\eqno{(2.13)}$$ In
the same manner the term $\Sigma|\mathcal{M}_{2}^{Z}|^{2}$ can be
evaluated to obtain an expression similar to equation (2.13). In
this case $P$ will be replaced by $Q$, where
$$Q=p_{1}-q$$
Evaluating the various trace terms rigorously and simplifying
those expressions we obtain
$$\int\sum|\mathcal{M}^{Z}|^{2}\frac{d^{3}q_{1}}{2q_{1}^{0}}\frac{d^{3}q_{2}}{2q_{2}^{0}}
\delta^{4}(q-q_{1}-q_{2})=F(p_{1}, p_{2}, p'_{1},
p'_{2})\eqno{(2.14)}$$ The right hand side of this equation is a
scalar obtained by the various combinations of the scalar product
of initial and final momenta of the electrons. Thus ultimately $F$
becomes the function of either energies or momenta of incoming and
outgoing electrons. The scattering cross section of this process
can be calculated by using the formula
$$\sigma=\frac{\mathcal{S}}{4\sqrt{(p_{1}p_{2})^{2}-m_{e}^{4}}}N_{p_{1}}N_{p_{2}}
\frac{1}{(2\pi)^{2}}\int
\frac{N_{p'_{1}}d^{3}p'_{1}}{2p_{1}^{'0}(2\pi)^{3}}
\frac{N_{p'_{2}}d^{3}p'_{2}}{2p_{2}^{'0}(2\pi)^{3}}
N_{q_{1}}N_{q_{2}}F(p_{1},p_{2},p'_{1},p'_{2})\eqno{(2.15)}$$ Here
all incoming and outgoing particles are spin-$\frac{1}{2}$
fermions. For that reason $N_{i}$
$(i=p_{1},p_{2},p'_{1},p'_{2},q_{1}, q_{2})$ is twice the mass of
the corresponding fermion. The square root term present in the
denominator comes from the incoming flux that is directly
proportional to the relative velocity of the incoming electrons
and written in Lorentz invariant way. As the final state contains
the incoming particle there must be a non-unit statistical
degeneracy factor $\mathcal{S}$ given by
$$\mathcal{S}=\prod_{l}\frac{1}{g_{l}!}$$ if there are $g_{l}$
particles of the kind $l$ in the final state. This factor arises
since for $g_{l}$ identical final particles there are exactly
$g_{l}!$ possibilities of arranging those particles; but only one
such arrangement is measured experimentally. We Calculate the
expression for $F(p_{1},p_{2},p'_{1},p'_{2})$ in the equation
(2.15) and then integrating that expression the scattering cross
section is obtained. We are interested to obtain a clear
analytical expression and so we do not use any numerical
technique. Instead, with some special choice of approximations we
have calculated the integral present in (2.15). Let us consider
the following four vector
$$p'=p_{1}'+p_{2}'=p_{1}+p_{2}-q$$
It is clear that $p'$ is timelike i.e. $(p')^{2}>0$ and so we can
take a proper Lorentz transform such that $p'=(p^{'0}, 0)$. To
obtain the integral over $d^{3}p'_{1}$ we should write
$$\int \frac{F}{p^{'0}_{2}} d^{3}p'_{2}=\frac{4\pi}{3}
\frac{\mid\overrightarrow{p}'_{1}\mid^{3}}{p^{'0}_{1}}
F(\mid\overrightarrow{p}'_{2}\mid=
\mid\overrightarrow{p}'_{1}\mid,...)+\epsilon$$ The error term
$\epsilon$ will be present since we consider the system in CM
frame, defined by the equation (2.6), simultaneously. Here
neglecting this error term we can use the following approximation.
$$\int \frac{F}{p^{'0}_{2}} d^{3}p'_{2}\approx\frac{4\pi}{3}
\frac{\mid\overrightarrow{p}'_{1}\mid^{3}}{p^{'0}_{1}}
F(\mid\overrightarrow{p}'_{2}\mid=
\mid\overrightarrow{p}'_{1}\mid,...)\eqno{(2.16)}$$ Next we
integrate over $d^{3}p'_{1}$ without any more approximation and
obtain the expression for the scattering cross-section as follows:
$$\sigma=\frac{(C_{V}^{2}+C_{A}^{2})}{9\pi^{2}}(\frac{eg}{M_{Z}cos\theta_{W}})^{4}
\frac{(p^{0})^{2}}{\sqrt{1-(\frac{m_{e}}{p^{0}})^{2}}}
[\ln(\frac{p^{0}}{m_{e}})+f(p^{0},r)]\eqno{(2.17)}$$ where,\\
$f(p^{0},r)=\ln r
-[r-\frac{m_{e}}{p^{0}}][14-\frac{16}{(1+\frac{C_{V}^{2}}{C_{A}^{2}})}
(\frac{m_{e}}{p^{0}})^{2}+\frac{3(1+\frac{3C_{V}^{2}}{2C_{A}^{2}})}
{(1+\frac{C_{V}^{2}}{C_{A}^{2}})}(\frac{m_{e}}{p^{0}})^{4}]
+\frac{1}{2}[r^{2}-(\frac{m_{e}}{p^{0}})^{2}]
[31-\frac{12(1+\frac{27C_{V}^{2}}{24C_{A}^{2}})}
{(1+\frac{C_{V}^{2}}{C_{A}^{2}})}(\frac{m_{e}}{p^{0}})^{2}-3(\frac{m_{e}}{p^{0}})^{4}]
-\frac{2}{3}[r^{3}-(\frac{m_{e}}{p^{0}})^{3}][7-3(\frac{m_{e}}{p^{0}})^{2}]
+\frac{1}{4}[r^{4}-(\frac{m_{e}}{p^{0}})^{4}]-3(\frac{m_{e}}{p^{0}})^{2}\ln(\frac{rp^{0}}{m_{e}})
[\frac{4(1+\frac{27C_{V}^{2}}{24C_{A}^{2}})}{(1+\frac{C_{V}^{2}}{C_{A}^{2}})}
-(\frac{m_{e}}{p^{0}})^{2}
-\frac{1}{(1+\frac{C_{V}^{2}}{C_{A}^{2}})}(\frac{m_{e}}{p^{0}})^{4}]
+3(\frac{m_{e}}{p^{0}})^{2}[\frac{p^{0}}{m_{e}}-\frac{1}{r}]
[2-\frac{(1+\frac{2C_{V}^{2}}{3C_{A}^{2}})}
{(1+\frac{C_{V}^{2}}{C_{A}^{2}})}(\frac{m_{e}}{p^{0}})^{2}]
-\frac{3}{2}(\frac{m_{e}}{p^{0}})^{4}[(\frac{p^{0}}{m_{e}})^{2}-(\frac{1}{r})^{2}]
[1-\frac{1}{(1+\frac{C_{V}^{2}}{C_{A}^{2}})}
(\frac{m_{e}}{p^{0}})^{2}]$\\
and
$$\frac{m_{e}}{p^{0}}<r=\frac{max(p^{'0}_{1},p^{'0}_{2})}{p^{0}}<1$$
$p^{0}$ represents the center of mass energy i.e.
$$p_{1}^{0}=p_{2}^{0}=p^{0}$$ whereas $p^{'0}_{1}$ and $p^{'0}_{2}$ stand
for energies of the outgoing electrons.\\\indent It is worth
noting that all three type of neutrinos may involve in this
process, since there is no lepton number violation for it. So far
we have used the technique applicable for both muon and tau
neutrino; but for electron type of neutrino another 8 Feynman
diagrams having $e-W^{-}-\nu_{e}$ effect may contribute in this
process. Four of them are for direct processes (Figure-3) and the
rest four represent exchange diagrams (Figure-4). These extra
diagrams contribute in the calculations of scattering
cross-section, but only for electron type of neutrino emission. In
that case the matrix element will be modified as
$$\mathcal{M}=\mathcal{M}^{Z}+\mathcal{M}^{W}\eqno{(2.18)}$$
where,
$$\mathcal{M}^{W}=-\frac{4\pi i e^{2}g^{2}}{8M_{W}^{2}}
[(\mathcal{M}_{1}^{W}+\mathcal{M}_{2}^{W}+\mathcal{M}_{3}^{W}+
\mathcal{M}_{4}^{W})-(\mathcal{M}_{5}^{W}+\mathcal{M}_{6}^{W}+\mathcal{M}_{7}^{W}+
\mathcal{M}_{8}^{W})]\eqno{(2.19)}$$
$$\mathcal{M}_{1}^{W}=[\overline{u}(p'_{1})(1-\gamma_{5})\gamma_{\rho}
\frac{(q^{\tau}\gamma_{\tau}+p_{1}^{'\tau}\gamma_{\tau}+m_{e})}
{(q+p'_{1})^{2}-m_{e}^{2}+i\epsilon}\gamma_{\mu}v_{\nu}(q_{2})]
[\overline{u}(p'_{2})\frac{\gamma^{\mu}}{(p_{2}-p'_{2})^{2}+i\epsilon}u(p_{2})]
[\overline{u}_{\nu}(q_{1})(1-\gamma_{5})\gamma^{\rho}u(p_{1})]\eqno{(2.20a)}$$
$$\mathcal{M}_{2}^{W}=[\overline{u}(p'_{1})\gamma_{\mu}
\frac{(-q^{\tau}\gamma_{\tau}+p_{1}^{\tau}\gamma_{\tau}+m_{e})}
{(q-p_{1})^{2}-m_{e}^{2}+i\epsilon}(1-\gamma_{5})\gamma_{\rho}v_{\nu}(q_{2})]
[\overline{u}(p'_{2})\frac{\gamma^{\mu}}{(p_{2}-p'_{2})^{2}+i\epsilon}u(p_{2})]
[\overline{u}_{\nu}(q_{1})(1-\gamma_{5})\gamma^{\rho}u(p_{1})]\eqno{(2.20b)}$$
Other $\mathcal{M}^{W}_{i}$'s (i=3,..8) have the similar
expressions as defined in the equations (2.4) and (2.5). We use
Fierz rearrangement to obtain the full expression for
$\mathcal{M}$ containing the contributions for both $Z$ and $W$
bosons exchanged diagrams. If we introduce Fierz rearrangement on
$\mathcal{M}_{1}^{W}$ in (2.20a) and add it to (2.2) we obtain
$$\mathcal{M}_{1}=[\overline{u}(p'_{1})(C_{V}'-C_{A}'\gamma_{5})\gamma_{\rho}
\frac{(q^{\tau}\gamma_{\tau}+p_{1}^{'\tau}\gamma_{\tau}+m_{e})}
{(q+p'_{1})^{2}-m_{e}^{2}+i\epsilon}\gamma_{\mu}u(p_{1})]
[\overline{u}(p'_{2})\frac{\gamma^{\mu}}{(p_{2}-p'_{2})^{2}+i\epsilon}u(p_{2})]
[\overline{u}_{\nu}(q_{1})(1-\gamma_{5})\gamma^{\rho}v_{\nu}(q_{2})]\eqno{(2.21a)}$$
and thus the total scattering matrix becomes
$$\mathcal{M}=-\frac{4\pi i e^{2}G_{F}}{\sqrt{2}}
[(\mathcal{M}_{1}+\mathcal{M}_{2}+\mathcal{M}_{3}+
\mathcal{M}_{4})-(\mathcal{M}_{5}+\mathcal{M}_{6}+\mathcal{M}_{7}+
\mathcal{M}_{8})]\eqno{(2.21b)}$$
where,$$C'_{V}=\frac{1}{2}+2sin^{2}\theta_{W}\hspace{2cm}
C'_{A}=-\frac{1}{2}$$ and
$$\frac{G_{F}}{\sqrt{2}}=\frac{g^{2}}{8M_{W}^{2}}=\frac{g^{2}}{8M_{Z}^{2}cos^{2}\theta_{W}}$$
Each $\mathcal{M}_{i}$ (i=1,2..8) is formed by adding
$\mathcal{M}^{Z}_{i}$ and rearranged $\mathcal{M}^{W}_{i}$. Then
we proceed in the same way as before and calculate the scattering
cross section as
$$\sigma_{\nu_{e}}=\frac{4(C_{V}^{'2}+C_{A}^{'2})}{9\pi^{2}}\alpha^{2}G_{F}^{2}
\frac{(p^{0})^{2}}{\sqrt{1-(\frac{m_{e}}{p^{0}})^{2}}}
[\ln(\frac{p^{0}}{m_{e}})+f_{\nu_{e}}(p^{0},r)]\eqno{(2.22)}$$ The
expression $f_{\nu_{e}}(p^{0},r)$ present in the equation (2.22)
is almost similar to the expression $f(p^{0},r)$. In fact if we
replace $C_{V}$ and $C_{A}$ present in $f(p^{0},r)$ by $C'_{V}$
and $C'_{A}$ respectively the expression $f_{\nu_{e}}(p^{0},r)$
can be found out. \\\indent The equation (2.22) gives the
scattering cross-section for electron type of neutrino whereas the
scattering cross section for both muon and tau neutrino can be
obtained by using equation (2.17). We like to find the scattering
cross section in the extreme relativistic as well as
non-relativistic limit. In those two limit the total scattering
cross section, in c.g.s unit, for all three type of neutrinos are
approximated as
$$\sigma_{\nu_{e},\nu_{\mu},\nu_{\tau}}\approx 5.8\times
10^{-50}\times(\frac{E_{ER}}{m_{e}c^{2}})^{2}
\ln(\frac{E_{ER}}{m_{e}c^{2}})\hspace{0.cm}cm^{2}\hspace{0.5cm}
[extreme-relativistic]\eqno{(2.23)}$$
$$\approx 3.44 \times
10^{-49}\times(\frac{E_{NR}}{m_{e}c^{2}})^{\frac{1}{2}}
\hspace{1.7cm}cm^{2}\hspace{0.5cm}[non-relativistic]\eqno{(2.24)}$$
To be noted that $E_{ER}$ and $E_{NR}$ represent the energy of the
single electron  related to extreme relativistic and
non-relativistic limit respectively.\\\indent We are going to
check the goodness of our approximated analytical method. To
verify it let us obtain the scattering cross-section for electron
type of neutrino in the relativistic case from the equation
(2.22), which gives
$$\sigma_{\nu_{e}}\simeq 4.16 \times
10^{-51}\times(\frac{E}{m_{e}c^{2}})^{2}
\ln(\frac{E}{2m_{e}c^{2}})\hspace{0.5cm}cm^{2}\eqno{(2.25)}$$
where $E$ is the CM energy.\\ In the Tabe-1 this result is
compared to the scattering cross section for electron type of
neutrino, obtained by using CalcHep software (version-2.3.7).
\section{Calculation of energy loss rate :}
A number of different cases are to be considered to calculate the
energy loss rate via electron neutrino bremsstrahlung process. We
have already obtained the scattering cross section in extreme
relativistic and non-relativistic limit. The later stage of the
stellar evolution may be degenerate as well as non-degenerate
characterized by the chemical potential. In the evolution of some
stars the electron gas exerts degenerate pressure which prevents
the star from contraction due to the gravitational force. The
complete degenerate gas is such in which all the lower states
below the Fermi energy become occupied. There may be some ranges
of temperature and density where electron energy would not be
bounded by Fermi-energy. Such non-degeneracy may be evident for
both relativistic and non-relativistic limit i.e. for $\kappa
T<m_{e}c^{2}$ and $\kappa T>m_{e}c^{2}$ respectively, where
$\kappa$ represents Boltzmann's constant. We calculate the energy
loss rate by using the formula
$$\rho\mathcal{E_{\nu}}=\frac{4}{(2\pi)^{6}\hbar^{6}}
\int_{0}^{\infty}\int_{0}^{\infty}\frac{d^{3}p_{1}}{[e^{\frac{E_{1}}{\kappa
T}-\psi}+1]}\frac{d^{3}p_{2}}{[e^{\frac{E_{2}}{\kappa
T}-\psi}+1]}(E_{1}+E_{2}) \int
d\sigma(E'_{1},E'_{2})g(E'_{1},E'_{2})|\overrightarrow{v_{1}}-\overrightarrow{v_{2}}|
\eqno{(3.1)}$$ where $\rho$ is the mass density of the electron
gas and $g(E'_{1},E'_{2})$ stands for Pauli's blocking factor,
given by
$$g(E'_{1},E'_{2})=[1-\frac{1}{e^{\frac{E'_{1}}{\kappa T}-\psi}+1}]
[1-\frac{1}{e^{\frac{E'_{2}}{\kappa T}-\psi}+1}]\eqno{(3.2a)}$$
$\psi=\frac{\mu}{\kappa T}$ ($\mu$ represents the chemical
potential of the electron gas) is related to the number density of
the electron by the following relation.
$$n=\frac{2(\kappa T)^{3}}{\pi^{2}(c\hbar)^{3}}
\int_{0}^{\infty}\frac{x[x^{2}-(\frac{m_{e}c^{2}}{\kappa
T})^{2}]^{\frac{1}{2}}} {[e^{x-\psi}+1]}dx\eqno{(3.2b)}$$ To be
noted that in equations (3.1) and (3.2b) the upper limit of the
momentum has been taken up to infinity; this may give an
impression that the CM energy of the electron may be very high and
comparable to $M_{Z}$ or $M_{W}$. It is not true since the
temperature and density of the stellar core in the later stage do
not allow the electron to gain that very high energy. Thus in the
equations (3.1) and (3.2b) the upper limit of the momentum depends
on the temperature and density of the electron gas. We go through
the different cases in the followings.\\Case I: In the extreme
relativistic non-degenerate case, i.e. when $T\gg 5.9\times
10^{9}$ K and $\rho < 2\times 10^{6}$ $gm/cc$, the chemical
potential becomes very small compared to $E_{ER}$. In this case
the energy loss rate is calculated as
$$\mathcal{E_{\nu}}\approx 5.04\times10^{12}\times T_{10}^{6}
[1+0.82\ln(1.7\hspace{0.1cm}T_{10})]\hspace{0.8cm}erg-gm^{-1}sec^{-1}
\eqno{(3.3)}$$ where $$T_{10}=T\times 10^{-10}$$ From this
analytical expression it is found that the energy loss rate does
not depend on the density of the electron gas so far the non-degeneracy remains effective.\\
Case-II: In the extreme relativistic degenerate region the density
would be very high; it is to be noted that for core temperature
$T\gg 5.9\times 10^{9}$ K and the density would be much higher
than $10^{7}$ $gm/cc$. Pauli's blocking factor plays an important
role to calculate the energy loss rate for degenerate electron.
For extreme relativistic case it can be approximated as
$$\int d\sigma(E'_{1},E'_{2})g(E'_{1},E'_{2})\approx e^{2(1-x_{F})}\sigma\eqno{(3.4a)}$$
where $x_{F}$ represents the ratio of the Fermi temperature to the
maximum temperature of the degenerate electron gas at the maximum
density ($\sim 10^{15}$ $gm/cc$). The energy loss rate in the
extreme degenerate case is obtained as
$$\mathcal{E_{\nu}}\approx6.56\times10^{10}\times T_{10}^{6}
[1+0.56\ln(1.7\hspace{0.1cm}T_{10})]\hspace{0.8cm}erg-gm^{-1}sec^{-1}
\eqno{(3.4b)}$$ To be noted that the above expression for energy
loss rate depends on density only, not on the temperature.\\
Case-III: The non-relativistic effect becomes important when the
central temperature of the star is below the $5. 9\times 10^{9}$ K
and the electron would be non-degenerate if
$(\frac{\rho}{2})^{\frac{2}{3}}\leq (\frac{T}{2. 97\times 10^{5}
K}$ ). In this case $\mu<m_{e}c^{2}$, but $\psi$ cannot be
neglected as in the extreme relativistic case. We calculate the
energy-loss rate as follows:
$$\mathcal{E_{\nu}}\approx0.88\times10^{-3}\times T_{8}\rho\hspace{0.8cm}
erg-gm^{-1}sec^{-1}\eqno{(3.5)}$$ Where $T_{8}$ is defined in the
same manner as $T_{10}$. This energy loss rate is not low in the
region having the temperature $10^{8}- 10^{9}$ K and density less
than $10^{6}$ gm/cc, which indicates the importance of this
process in the non-relativistic non-degenerate region.\\\indent It
is worth noting that the energy loss rate in the non-relativistic
but degenerate case is very low, hence this case is not considered
here.
\section{Discussion :} In the calculations of scattering cross-section we have used an
approximation in the equation (2.16), but the Table-1 shows our
result is very close to that generated by the software. It
strongly supports the approximation we have used in the equation
(2.16). The scattering cross section obtained under the frame-work
of electro-weak theory is very small, especially in the
non-relativistic case, but as the mean free path of the neutrinos
is much longer than the scale of stellar radius the
electron-neutrino bremsstrahlung process may have some effect to
release energy from star at high temperature and density. The
relativistic effect comes into play when the temperature exceeds
$6\times 10^{9}$ K. It is evident from our work that the
electron-neutrino bremsstrahlung process yields a large amount of
energy loss from the stellar core when core temperature $\geq
10^{10}$ K, both in non-degenerate as well as degenerate region.
In that temperature range the radiation pressure is so dominating
that the gas pressure has negligible effect \cite{Chandrasekhar}.
In this extreme relativistic region the process contributes
significantly when the electron gas is non-degenerate. That was
also shown by Cazzola and Saggion \cite{Cazzola}. But they did not
calculate the energy loss rate in the degenerate region; though
they indicated that the electron neutrino bremsstrahlung process
might be highly significant in that region. We have calculated the
energy-loss rate to obtain an analytical expression for the energy
loss rate when the electrons are strongly degenerate. A typical
example of the extreme-relativistic degenerate stellar object is
the newly born neutron star, which is the result of type-II
Supernova. Our study reveals that the energy loss rate in the
non-degenerate region is higher than that calculated in the
degenerate region. This clearly indicates that though during the
neutron star cooling electron-neutrino bremsstrahlung plays a
significant role, but the process becomes more important to carry
away the energy from the core of pre-Supernova star, which is a
relativistic non-degenerate stellar object.\\\indent
Non-relativistically the process becomes insignificant unless the
temperature is sufficiently high, at least the temperature should
attain $10^{8}$ K. At this temperature the burning of helium gas
in the stellar core takes place \cite{Hayashi}. In the temperature
range $10^{8}- 10^{9}$ K the gas pressure is dominating over the
radiation pressure; though the effect of radiation pressure cannot
be neglected in this region. In addition, the region will be
non-degenerate if density $<2\times 10^{6}$ gm/cc. The electron
neutrino bremsstrahlung process may have some effect in this
region though the energy loss rate is not so high as it is in the
extreme relativistic case. In the low density the energy loss rate
increases rapidly with rising core temperature. Eventually the
process contributes in both degenerate and non-degenerate cases,
whereas degenerate electrons participate only when the density of
the medium is very high.  Hence, the electron-neutrino
bremsstrahlung process is an important energy-generation mechanism
during the evolution of stars, particularly in the later stages.
\section{Acknowledgement:}
It is my pleasure to thank Prof. {\bf Probhas Raychaudhri},
University of Calcutta, India for his valuable suggestions and
guidance in the preparation of this manuscript. I would also like
to thank {\bf CSIR}, India for funding this research work and {\bf
ICTP}, Trieste, Italy for providing me some excellent facilities
and information, which have facilitated to carry out this work. My
special thank goes to {\bf Biplob Bhattacherjee}, Department of
Pure Physics, University of Calcutta for helping me in the
computational work.

\pagebreak
\begin{table}
\noindent{\large\bf  \quad Table-1 :} \vspace{0.5cm}\\
\begin{tabular}{|c|cc|}\hline
CM Energy $(MeV)$&\hspace{3cm}$\sigma_{\nu_{e}}(cm^{2})$\\\hline &
Software generated result  & Our result \\\hline
  10 & $3.45\times 10^{-48}$ & $3.64\times 10^{-48}$ \\
  20 & $1.19\times 10^{-47}$ & $1.88\times 10^{-47}$ \\
  30 & $4.47\times 10^{-47}$ & $4.84\times 10^{-47}$ \\
  40 & $1.17\times 10^{-46}$ & $0.92\times 10^{-46}$ \\
  50 & $1.23\times 10^{-46}$ & $1.56\times 10^{-46}$ \\
  60 & $1.95\times 10^{-46}$ & $2.32\times 10^{-46}$\\
  70 & $4.06\times 10^{-46}$ & $3.28\times 10^{-46}$ \\
  80 & $4.45\times 10^{-46}$ & $4.44\times 10^{-46}$ \\\hline
\end{tabular}
\caption{Comparison of the scattering cross-section for electron
type of neutrino obtained by our method relative to that generated
by CalcHep software.}
\end{table}
\vspace{0.5cm} \noindent{\large\bf \quad Figure Caption :}
\vspace{0.2cm}\\ Figure-1: Feynman diagrams for the direct process\\
Figure-2: Exchange diagrams\\
Figure-3: Feynman diagrams for the direct
process having $e-W^{-}-\nu_{e}$ effect\\
Figure-4: Exchange diagrams having $e-W^{-}-\nu_{e}$ effect
\end{document}